\documentclass[6pt]{amsart}
\usepackage{graphicx}
\usepackage{moreverb}
\usepackage{booktabs}
\usepackage{placeins}
\usepackage{natbib}        
\usepackage{amsmath} 
\usepackage{mathrsfs}
\usepackage[colorlinks=true, citecolor=black, linkcolor=black, urlcolor=black]{hyperref} 
\usepackage{tikz, tikz-3dplot}
\usetikzlibrary{calc,angles,quotes}
\tdplotsetmaincoords{70}{110}
\usepackage{bbm}
\usepackage{thm-restate}
\usepackage{bm}
\usepackage{natbib} 
\usepackage{booktabs,tabularx, colortbl}
\usepackage{amsmath}
\usepackage{mathtools}
\usepackage{dsfont}
\usepackage{textcomp}
\usepackage{xcolor}
\usepackage[english]{babel}
\newcommand{\RNum}[1]{\uppercase\expandafter{\romannumeral #1\relax}}

\DeclareMathOperator{\E}{\mathbb{E}}

\newcommand{\eps}{\varepsilon}

\newcommand{\X}{\mathbb{X}}

\usepackage{xargs}
\newcommandx{\pto}[0]{\overset{P}{\to}}

\newcommand\BibTeX{{\rmfamily B\kern-.05em \textsc{i\kern-.025em b}\kern-.08em
T\kern-.1667em\lower.7ex\hbox{E}\kern-.125emX}}

\usepackage{indentfirst,csquotes}

\usepackage{amssymb,amsthm,amsmath}
\usepackage{xcolor,paralist,hyperref,fancyhdr,etoolbox}

\hypersetup{ colorlinks=true, linkcolor=black, filecolor=black, urlcolor=black }

\usepackage{lipsum}

\begin{document}
\title{``Within-trial'' prognostic score adjustment is targeted maximum likelihood estimation} 
\author{Emilie Højbjerre-Frandsen*}

\author{Alejandro Schuler}

\date{\today}


\let\thefootnote\relax

\begin{abstract}
Adjustment for ``super'' or ``prognostic'' composite covariates has become more popular in randomized trials recently. These prognostic covariates are often constructed from historical data obtained from previous clinical trials or registries by fitting a predictive model of the outcome on the raw covariates. A natural question that we have been asked by applied researchers is whether this can be done without the historical data: can the prognostic covariate be constructed or derived from the trial data itself, possibly using different folds of the data, before adjusting for it? Here we clarify that such ``within-trial'' prognostic adjustment is nothing more than a form of targeted maximum likelihood estimation (TMLE), a well-studied procedure for optimal inference. \textcolor{black}{We therefore argue that there is no reason to reinvent the wheel, and that within-trial prognostic score adjutstment should just be refered to as a TMLE}. We \textcolor{black}{finally} discuss the pros and cons of within-trial prognostic adjustment (standard efficient estimation) relative to standard prognostic adjustment with historical data.
\end{abstract} 

\maketitle

\markright{``Within-trial'' prognostic score adjustment is targeted maximum likelihood estimation}

\section{Introduction}
Recently, there has been an increase in the exploration of "super covariates" or "prognostic covariates"  \citep{Holzhauer2022} as effective tools for covariate adjustment in statistical analysis in randomized clinical trials (RCTs). \textcolor{black}{The idea, which in principle dates back at least to \citet{tukey1993tightening}, is to utilize historical data to develop a single prognostic covariate, facilitating an increase in efficiency with reduced risk of overfitting}. This strategy effectively combines insights from earlier trials to strengthen the overall study power \textcolor{black}{\citep{CUPED, Schuler2020, tukey1993tightening, Holzhauer2022}}. \citet{Schuler2020}, formalized some of these ideas into the PROCOVA\texttrademark\footnote{The authors of \citet{Schuler2020} were all employed at \href{https://www.unlearn.ai}{Unlearn.AI} at the time of publication.} approach, which allows for adjustment of a prognostic score in a linear analysis \textcolor{black}{and translated the potential power gains to sample size gains}. We will refer to this method as \textcolor{black}{standard} prognostic score adjustment. This method preserves the randomization process and mitigates the increase in type I errors seen for other data fusion methods, \citep{Lim2018, hill, kaplan}. \citet{Schuler2020} demonstrated that the estimate of the average treatment effect (ATE) derived from this method is efficient, given the assumption of homogeneous treatment effects. \citet{HojbjerreFrandsen2024} examine the practical aspects associated with the application of this method. 

Another example of how to increase efficiency by adjusting for baseline covariates is Targeted Maximum Likelihood Estimation (TMLE) \citep{vanderLaanRubin+2006, intro_modern_CI}. The method is suitable for a wide range of data types and does not require the availability of historical data. There are numerous extensions that utilize cross-validation to enhance efficiency \citep{vanderLaanGruber+2010, GrubervanderLaan+2010, Weixin+vdLaan2020, chen2023, RosenblumvanderLaan+2010, Balzer2016, Balzer2024}. In many instances, the estimator achieves local semi-parametric efficiency, with discussions by \citep{intro_modern_CI, vanderLaanRubin+2006, vanderLaan+2017} outlining the requirements for TMLE to reach non-parametric efficiency. Additionally, \citet{Liao2023} expanded the method of prognostic score adjustment to encompass estimators that are already efficient, which is usually the case for TMLE under an RCT.

In this \textcolor{black}{paper} we explore ``within-trial'' prognostic score adjustment. This method leverages the ideas of prognostic score adjustment, but eliminates the need for historical data. \textcolor{black}{We start by setting the theoretical framework for ATE estimation in an RCT. We next describe how a linear model can be used for ATE estimation, following how the efficiency of this procedure can be increased using prognostic scores. We next explain TMLE for ATE estimation in a 1:1 RCT setting and use this to} argue that within-trial prognostic adjustment is a type of TMLE for ATE estimation in RCTs. \textcolor{black}{Finally, we discuss the impact of the equivalence between the two methods, and compare this to the standard prognostic score adjustment proposed by \citet{Schuler2020}.}

\section{Setting and notation}

Our setting is a two-arm trial involving $n$ participants $O_i = \left(W_i, A_i, Y_i\right)$, which are independent and identically distributed. We refer to a generic observation as $O=\left (W,A,Y\right)$ without index $i$. Here $Y$ represents a continuous primary endpoint variable and $W$ is a vector of $p$ baseline covariates. $A$ is the treatment indicator, which is $1$ if the participant is randomized to the new treatment and $0$ otherwise. We do not impose any parametric assumptions on the distribution of $Y$ given $(A, W)$. The trial dataset is represented as $\left(\mathbb{W}, \mathbb{A}, \mathbb{Y}\right)\in \mathcal{W}^n\times \{0, 1\}^n \times \mathbb{R}^n$, where $\mathcal{W}$ is the domain of $W$. The sizes of the treatment and control groups are represented as $n_1$ and $n_0$, respectively. 

Each participant is associated with two potential outcomes: $Y(1)$ under the new treatment and $Y(0)$ under control \citep{Petersen+vdLaan2014, Sekhon2008, Imbens2004}. The estimand of interest is the causal average treatment effect (ATE), defined mathematically as:
\begin{align}\label{eq:causalestimand}
    \begin{split}
        \Psi^* = \E [Y(1)-Y(0)].
    \end{split}
\end{align}
\textcolor{black}{Without further assumptions, there is no way to estimate this causal effect because we can not observe both potential outcomes simultaneously, even with infinite draws from the $(Y, A, W)$ data structure. However, there are multiple sets of assumptions that one can impose on the causal data generating process $(Y(1),  Y(0), A, W)$ so that $\E[Y(a)]$ is equal to some parameter (functional) of the observed data distribution, assuming that the observed outcome is $Y = Y(A)$ (whichever potential outcome corresponds to observed treatment).}

\textcolor{black}{The most parsimonious of these is assuming simple randomization. Due to randomization, the potential outcomes $Y(0)$ and $Y(1)$ are independent of the treatment allocation $A$ and the exposure probabilities $1>P\left(A = a \right) = \pi_a>0$ (for $a\in \{0,1\}$) are known. In this case the identifying functional becomes $\E[Y(A)] = \E[\E[Y|A,W]]$. In much of the literature, this is called the \textit{statistical estimand} (in contrast to the \textit{causal estimand}). The two are identical under the identifying assumptions but the form that the statistical estimand takes depends on what the identifying assumptions are. A well-known benefit of identification via randomization is that the randomization assumption can be enforced by design, as is done in a randomized trial.} Thus the causal estimand \eqref{eq:causalestimand} is identified by a corresponding statistical estimand:
\begin{align}\label{eq:statestimand}
    \begin{split}
        \Psi&= \E [ Y(1)]-\E [Y(0)] = \E\big[ \E[ Y|A=1, W]-\E[Y|A=0, W]\big]\\
        &= \E[\mu\left(1, W\right)-\mu\left(0, W\right)] ,
    \end{split}
\end{align}
where $\mu(a, W)=\E[ Y|A=a, W]$ is the conditional mean function. \textcolor{black}{In the remainder of the paper we treat the identifying functional (statistical estimand) $\Psi$ as the target parameter and focus on estimators and inference for this under simple randomization.}

\section{ATE estimation methods in RCTs}\label{sec:ateestimation}

\textcolor{black}{In this section we explain three approaches to estimate $\Psi$ in RCTs: the standard ANCOVA (plug-in) estimator, prognostic score adjustment using historical data, and TMLE. These definitions are used to set up the equivalence result in Section~\ref{sec:within}.}

\subsection{ANCOVA Estimator}\label{sec:plug-in_with}

A simple linear model including an intercept and a treatment term is often used to estimate the ATE for a continuous endpoint in the pharmaceutical industry. \textcolor{black}{In such a model the mean vector for the endpoint $\mathbb{Y}$ is modelled as $\beta_01_n+\mathbb{X}\beta$ where $1_n$ is a vector of ones and $\mathbb{X}$ is the design matrix including a column consisting of the treatment indicator, $\mathbb{A}$, as well as any covariates, interactions, etc. Maximum likelihood (MLE) is used to estimate $\beta_0$ and $\beta$. When the design matrix only includes the treatment indicator the estimator is known as the unadjusted estimator, and if the design matrix includes covariates the estimator is known as Analysis of Covariance (ANCOVA).}

If the design matrix includes no \textcolor{black}{treatment--covariate} interaction terms, or if the covariates are centered, then the ATE can be consistently estimated in an RCT as $\hat\Psi = \hat\beta_A$, the coefficient on the treatment indicator. The robust (\textcolor{black}{Huber--White}) standard error may typically be used for inference if some care is taken with centering the covariates \cite{MacKinnon1985, Long2000, FDA_cov, ye2021betterpracticecovariateadjustment, HojbjerreFrandsen2024}.

\textcolor{black}{While this is the standard presentation of ANCOVA, here we give a complementary account following \citet{RosenblumvanderLaan+2010} that is (under weak conditions) algebraically equivalent but reveals the close connection between ANCOVA, G-computation, and TMLE. 
In this account, the ANCOVA estimate is obtained by the following plug-in procedure:} 
\begin{enumerate}
    \item \textcolor{black}{Use MLE to estimate $\beta_0$ and $\beta$.}
    \item Extract the counterfactual predictions using the estimated conditional mean functions $\hat{\mu}(a,w)=\hat{\beta}_0+x\hat{\beta}_x$, i.e.
    \begin{align}
    \hat{\Psi}_a = \frac{1}{n}\sum_{i=1}^n \hat{\mu}(a, W_i),  \quad\quad  a \in \{0, 1\}. 
    \end{align}
    \item Get the ATE estimate by plugging in $\hat{\Psi} = \hat\Psi_1 - \hat{\Psi}_0$.
    \item The sampling variance of the influence function for this estimator gives a consistent estimate of the asymptotic variance,
    \begin{align}\label{eq:asympvar}
    \hat{\sigma}_\infty^2 = \frac{1}{n}\sum_{i=1}^n \left( \hat\phi_{1}(W_i, A_i, Y_i)  - \hat\phi_{0}(W_i, A_i, Y_i) \right)^2,
    \end{align}
where $\hat\phi_{a}(W_i, A_i, Y_i) = \dfrac{1_a(A_i)}{\pi_a}(Y-\hat\mu(a, W_i))+(\hat\mu(a, W_i)-\hat\Psi_a)$. Confidence intervals and p-values may then be constructed from the approximate standard error $\hat\sigma = \hat\sigma_\infty / \sqrt{n}$.
\end{enumerate}

\textcolor{black}{This procedure is a specific case of G-computation in the context of an RCT using a linear regression model as the estimate of the conditional mean of the outcome. This procedure is generally valid and avoids gotchas from interactions and centering while otherwise being algebraically equivalent to taking the coefficient on treatment $\hat\beta_A$ and constructing a robust standard error.}

\subsection{Prognostic Adjustment}

\textcolor{black}{
The plug-in formulation of the ANCOVA estimator illustrates that good estimation of the ATE depends on good estimation of the conditional mean $\hat\mu$. Indeed, adjustment works by explaining away variability in the outcome with variability in the covariates, leaving the effect of treatment more clearly visible. Therefore it is natural to try and get a better predictive model $\hat\mu$, which is often possible if there is historical data available.
}

\textcolor{black}{
The idea of prognostic adjustment is to improve the estimate of $\hat\mu$ with historical data while protecting from possible type I error inflation. Therefore, instead of directly influencing the outcome regression $\hat\mu$, the external data are used to learn a separate regression model $\hat\rho(w,a)$, the \textit{prognostic score}. This model is used on the trial data to generate predictions, which are then included as a ``prognostic'' covariate in a standard ANCOVA analysis. From the perspective of the ANCOVA, the covariate $\hat\rho(A,W)$ is a prespecified, fixed transformation of the predictors and therefore cannot affect type I error. However, if the external and trial populations are similar, it is likely to be a powerful and predictive summary measure which will improve estimation of $\hat\mu$ and therefore increase power for the ATE analysis.
}

Formally, to define \textcolor{black}{a} prognostic score we introduce the stochastic variable $D$, which is $1$ if the participant comes from the new trial and $0$ if the observation is from the historical data. The prognostic score is defined as the expected observed outcome conditional on the covariates
\begin{align}
    \begin{split}
        \rho_D\left(W, A\right) := \E[Y \,|\, W, A, D].
    \end{split}
\end{align}
This true expected value is called the oracle prognostic score. The trial-specific conditional mean which we are interested in to estimate the ATE can be expressed as $\mu(A,W) = \rho_1(\textcolor{black}{W, A})$. Thus if the external ($D=0$) population is similar to the trial population, $\rho_0 \approx \rho_1$ and we can leverage the external data to improve estimation of $\mu$. 

\textcolor{black}{Having access to historical data $\left(\widetilde{\mathbb{W}}, \widetilde{\mathbb{A}}, \widetilde{\mathbb{Y}}\right)\in \mathcal{W}^{\widetilde{n}} \times \{0,1\}^{\widetilde{n}} \times \mathbb{R}^{\widetilde{n}}$ obtained for $\widetilde{n}$ participants, estimators of the prognostic scores can be obtained by applying a machine learning algorithm to this data. Usually, there are no treatment units in the historical data meaning that $A=0$ for all participants and only one estimator $\widehat \rho_0(W, 0)$ of the prognostic score would be obtained.} In the standard prognostic score adjustment procedure proposed by \citet{Schuler2020} \textcolor{black}{this} prediction is then used as an additional adjustment covariate in the ANCOVA method, increasing predictive accuracy and thus efficiency. \textcolor{black}{As earlier stated \citet{Schuler2020} show that the ATE estimate is then efficient, given the assumption of homogeneous treatment effects. However, if the historical data does include treated participants one can estimate the prognostic scores as $\widehat \rho_0(W, A)$. In this case \citet{Schuler2020} has an equivalent result showing that the ATE estimate is now semi-parametrically efficient.} 

The \textcolor{black}{finite sample} gain in efficiency depends on the machine learning model's ability to capture non-nonlinearities and interaction effects that we were otherwise not able to capture through the linear model or learn adequately from a (usually smaller) trial dataset. \textcolor{black}{However, the purpose of this paper is not to argue why or how prognostic score adjustment increases efficiency. For exact statements on the requirements for efficiency gain see \cite{Schuler2020, Schuler+2022+151+171, tsiatis2007semiparametric, HojbjerreFrandsen2024}.}

\subsection{TMLE}\label{sec:tmle}

\textcolor{black}{
Instead of leveraging historical data via an ANCOVA model to improve estimation of the conditional mean $\hat\mu$, TMLE takes the more direct approach of using the trial data in combination with a more flexible and possibly more accurate estimator of $\hat\mu$. In general, TMLE uses estimates $\hat\mu$ from machine learning models instead of linear regression. However, this creates another problem that must be solved.
}

Recall that for the ATE we are trying to estimate $\E[\mu\left(1, W\right)-\mu\left(0, W\right)]$. If we could generate accurate estimates $\hat\mu \approx \mu$, via some kind of regression, a sensible estimator would be the ``plug-in'' $\frac{1}{n} \sum_i \hat\mu(1,W_i) - \hat\mu(0,W_i)$, the difference of the ``imputed'' potential means for each subject. The accuracy of that estimate is dependent, however, on the degree to which $\hat\mu \approx \mu$. It therefore behooves us to generate $\hat\mu$ using powerful machine learning methods that can capture any nonlinearities, interactions, etc. that are present in $\mu$. The efficiency gains from TMLE (and similar strategies like double machine learning) come from using more accurate regression models. 

Unfortunately, if $\hat\mu$ comes from a generic machine learning model, the plug-in estimator is generally too biased to allow for the construction of standard confidence intervals and p-values. This happens because $\hat\mu$ is optimized (e.g. via regularization and cross-validation model selection) to minimize some kind of predictive metric like mean squared error, which is a combination of bias and variance. Therefore $\hat\mu$ is typically biased even when machine learning is used and the bias does not reduce quickly enough with sample size to allow for standard inference. \textcolor{black}{For rigorous statements on the plug-in bias (using the von Mises expansion) one can look at \cite{intro_modern_CI, tsiatis2007semiparametric, kennedy2022semiparametric}, but we give an intuitive explanation below.}

As a concrete example, consider fitting a lasso or ridge model for $\hat\mu(a,w) = \alpha_0 + a\alpha_A + x\alpha$ in a randomized trial and imagine that the terms $x$ of the design matrix contain enough covariates and interactions that their linear combination can accurately model $\mu(a,w)$ in the population. Due to the presumably large number of terms in the design matrix, regularization is required and cross-validation will be necessary to choose the appropriate level. If the design matrix is centered, the plug-in estimate of the treatment effect will be $\frac{1}{n} \sum\hat\mu(1,W_i) - \hat\mu(0,W_i) = \hat\alpha_A$, the coefficient on treatment. But, due to the regularization, $\hat\alpha_A$ will be shrunken (biased downwards) relative to the population value. Another way to think of this is that the graphs of the two curves $w\mapsto \hat\mu(1,w)$ and $w\mapsto \hat\mu(0,w)$ will be closer to each other than they should be. 

TMLE solves this problem by ``updating'' the outcome prediction function $\hat\mu$ to a new function $\hat\mu^*$. The update step is done in such a way that $\hat\mu^*$ is debiased for purposes of a plug-in estimator. This can be thought of as a (targeted) ``relaxation'' of the implicit or explicit regularization used to fit $\hat\mu$. As a result, TMLE allows us to collect the efficiency gains from using nonparametric regression models without paying any price in bias.

In the context of a simple trial with 1:1 randomization there is a version of the TMLE update step is relatively easy to explain. The condition that any regression model $\tilde\mu$ would have to meet in order to be unbiased when used in a plug-in estimator of the ATE in an RCT is
\begin{align}
\E[\tilde\mu(1,W) - \tilde\mu(0,W)] 
&= \E[Y(1) - Y(0)] = \E[Y|A=1] - \E[Y|A=0]
\end{align}
because by randomization $\E[Y(a)] = \E[Y|A=a]$. Since we cannot take exact expectations in the sample to calculate $\E[Y|A=1] - \E[Y|A=0]$ we replace these terms with sample averages 
$$
\tilde\Psi = \underbrace{\frac{1}{n_1} \sum_{A_i=1} Y_i}_{\tilde\Psi_1} - \underbrace{\frac{1}{n_0} \sum_{A_i=1} Y_i}_{\tilde\Psi_0}
$$ 
This is just the unadjusted estimate of the effect, which is known to be unbiased \textcolor{black}{specifically} in an RCT \textcolor{black}{setting} \textcolor{black}{(but not in an observational study).} In other words, \textcolor{black}{in an RCT,} we would like our TML estimate to be equal on average to the unadjusted estimate. This seems redundant but it is important because we will use the updated regression models to perform \textit{inference} as well as to estimate the effect: that is where the gain in efficiency will come from. The point estimate will be the same (on average) as the unadjusted one, but the confidence intervals will generally be tighter because we will use the updated regression models to calculate the SE and reduce outcome variance that is explainable with covariates. The SE for TMLE is calculated the same way as shown in \autoref{eq:asympvar} for the ANCOVA plug-in estimator, but using the updated machine-learning regression $\hat\mu^*$, \textcolor{black}{i.e} 
\begin{align}\label{eq:var_tmle}
    \textcolor{black}{\hat{\sigma}_\infty^2 = \frac{1}{n}\sum_{i=1}^n \left( \dfrac{1_1(A_i)}{\pi_a}(Y-\hat\mu^*(1, W_i))+(\hat\mu^*(1, W_i)-\hat\Psi_1)  - \dfrac{1_0(A_i)}{\pi_0}(Y-\hat\mu^*(0, W_i))+(\hat\mu^*(0, W_i)-\hat\Psi_0) \right)^2.}
    \end{align}
\textcolor{black}{\citet{Schuler+2022+151+171} derive an expression for this variance (see their Appendix A.2) that makes it clear that the variance depends critically on the mean-squared error of $\hat\mu^*$}. The (usually larger) standard error for the unadjusted estimate can also be calculated with the same formula, setting $\hat\mu(a, W)$ equal to the constants $ \tilde\Psi_a$. 

\textcolor{black}{We have now motivated \textit{why} we would like to update the outcome regression (to debias it) but we have not explained \textit{how} to do so.}
Define $A_\pm = 2A-1$ and $a_\pm = 2a-1$, the treatment indicator mapped to $\{-1,1\}$. If we update the initial regression model using a single term
\begin{align}\label{eq:eps}
\begin{split}
\hat\mu^*(a,w) &= \hat\mu(a, w) + \epsilon^*a_\pm \\
\epsilon^* &= \frac{1}{n}(n_1\tilde\Psi_1 - n_0\tilde\Psi_0) - \frac{1}{n} \left(\sum_{A_i=1} \textcolor{black}{\hat\mu}(1, W) - \sum_{A_i=0} \textcolor{black}{\hat\mu}(0, W) \right)
\end{split}
\end{align}
then one may algebraically verify that $\E[\hat\mu^*(1,W) - \hat\mu^*(0,W)] = \tilde\Psi$ as desired when $n_1 = n_0 = n/2$.

Recall that our problem was that $\hat\mu(0,w)$ and $\hat\mu(1,w)$ are generally too close together due to regularization. The effect of adding the update $\epsilon^*(2a-1)$ is to push the two curves apart by an amount $2\epsilon^*$ while otherwise leaving them untouched. The quantity $\epsilon^*$ is engineered so that the resulting average distance between the two curves is exactly $\tilde\Psi$ as desired. 

It is no coincidence that $\epsilon^*$ also happens to be the empirical maximum likelihood solution to the estimation problem $Y = \hat\mu(A, W) + \epsilon A_\pm + \mathcal N(0,1)$ (verifying this again takes some algebra, \textcolor{black}{see Appendix~\ref{app1}}). Until now, our explanation of TMLE has been informal, but this observation allows us to generalize and formalize to some extent. The model which we are searching with maximum likelihood in the described update is the set of conditional densities 
$$
\{p_\epsilon(Y|A, W) \sim \mathcal N(\hat\mu(A, W) + \epsilon A_\pm, 1): \epsilon \in \mathbb R\}.
$$
The fact that our updated regression is the MLE in this model is why the method is called targeted \textit{maximum likelihood} \textcolor{black}{(we will explain the ``targeted'' part shortly)}. \textcolor{black}{We are doing maximum likelihood estimation in the ``targeting'' submodel shown above}. In the context of such a model, let $\E_\epsilon[\cdot] = \E_{p_{\epsilon}}[\cdot]$. The model shown above has the special properties that 
\begin{enumerate}
    \item the conditional mean at some value $\epsilon_0$ (in this case $\epsilon_0=0$) is $\E_{\epsilon_0}[Y|A, W] = \hat\mu(A, W)$, the initial regression fit, and
    \item the derivative of the log-likelihood in $\epsilon$ is given by $\frac{d}{d \epsilon} \log p_\epsilon = A_\pm(Y-\mu_\epsilon(A, W))$ where $\mu_\epsilon(A, W) = \E_\epsilon[Y|A, W]$.
\end{enumerate}
These properties are what make the model \textit{targeted} for the debiasing of the ATE in an RCT. 

\textcolor{black}{There is not one single definition of what a TMLE is. However, in the context of an 1:1 RCT} it has been shown that updating in \textit{any} parametric model $\{p_\epsilon(Y|A, W) : \epsilon \in \mathbb R^d\}$ that satisfies these two conditions will suffice for the TMLE update step \textcolor{black}{in the sense of providing asymptotic semiparametric efficiency}: one sets up the chosen model, finds the maximum likelihood estimate $\epsilon^*$ within it, and then takes $\hat\mu^*(A, W) = \E_{\epsilon^*}[Y|A, W]$ as the update. \textcolor{black}{In this paper we are in the setting of a 1:1 RCT, and we will therefore consider this as our formal definition of TMLE for ATE estimation. There are more definitions of TMLEs in different contexts and for different estimands \cite{vanderLaanRubin+2006, van2011targeted}.} 

\textcolor{black}{Updating in a parametric model that satisfies these two conditions will make} the resulting estimator be debiased the same way that our example was (using the additive normal update model) \citep{vanderLaanRubin+2006, intro_modern_CI}. When $\epsilon = [\epsilon_1, \epsilon_2, \dots \epsilon_d]$ is a multidimensional parameter, the second condition generalizes to $A_\pm(Y-\hat\mu(A, W))$ being contained in the linear span of the functions $\frac{\partial}{\partial \epsilon_j} \log p_\epsilon$. This happens in particular when there is some $\epsilon_j$ for which $\frac{\partial}{\partial \epsilon_j} \log p_\epsilon = A_\pm(Y-\mu_\epsilon(A, W))$.

\textcolor{black}{To summarize, TMLE constructs an updated outcome regression $\hat\mu^*$ by maximizing the likelihood within a submodel chosen so that the update removes the bias relevant to the plug-in using $\hat\mu$ without targeting. We can summarize the procedure for ATE estimation in an RCT as:}
\textcolor{black}{
\begin{enumerate}
    \item Obtain an initial estimator $\hat{\mu}$ for the conditional mean functions using a working outcome model, potentially some powerful machine learning method.
    \item From $\hat{\mu}$ obtain the predicted outcomes $\hat{\mu}(A,W)$ for each participant in the study. 
    \item Use MLE with the trial data and predictions to estimate $\epsilon^*$ using the regression model $Y = \hat\mu(A, W) + \epsilon A_\pm + \mathcal N(0,1)$.
    \item Use $\epsilon^*$ to get the targeted predictions $\hat\mu^*(0,w) = \hat\mu(0, w) - \epsilon^*$ and $\hat\mu^*(1,w) = \hat\mu(1, w) + \epsilon^*$ for each participant in the study. 
    \item Average the targeted predictions to estimate the counterfactual mean $\hat{\Psi}^\star_a$ for $a\in\{0, 1\}$. 
    \item Contrast on the ATE scale, i.e. $\hat{\Psi}^\star_1- \hat{\Psi}^\star_0$ and obtain the variance estimate using \eqref{eq:var_tmle}.
\end{enumerate}
}

\textcolor{black}{
A final wrinkle is that the initial outcome regression $\hat\mu$ generally needs to be ``cross-fit'' in order to assure $\sqrt n$-consistency of the point estimate and consistency of the standard error. This has been discussed elsewhere and is not essential to our presentation so we postpone any further discussion to Section~\ref{sec:discussion} \cite{cv_tmle, dml, aipw-tmle-2, aipw-tmle-3}.
}

\textcolor{black}{TMLE generally performs better than G-computation due to the debiasing. In the context of an RCT, inverse propensity weighting (IPW) is the same as the unadjusted estimate, which is less efficient than TMLE. Augmented IPW (AIPW) is another tool for debiasing the plug-in based on an estimate $\hat\mu$ from a machine learning model. AIPW is asymptotically equivalent to TMLE and they tend to behave very similarly. Many papers have examined the relationship between these different estimators, but this is not the focus of our current work \cite{aipw-tmle-1, aipw-tmle-2, aipw-tmle-3}.
}

\section{Within-trial prognostic score adjustment}\label{sec:within}

\textcolor{black}{Next we show that constructing a prognostic score within the trial and using it in an ANCOVA model is equivalent to performing a TMLE update in an RCT.}

\textcolor{black}{Clinical} trials are usually small, \textcolor{black}{so} using large external datasets to learn a prognostic score can be a good idea \cite{Schuler2020, Liao2023, HojbjerreFrandsen2024}. However, if external data are not available or trustworthy, one could imagine building the prognostic model on the trial data itself to eliminate the use of historical data. Specifically, one would fit a prognostic model on the trial data by regressing $Y$ onto $(A,W)$ with machine learning methods and then predict the outcome for each participant $\widehat \rho_1(A, W) = \hat\mu (A,W)$, now predicting the outcome using the actual treatment allocation. The idea is then to add the predictions as an additional covariate in the design matrix for the ANCOVA method, following the prognostic score adjustment procedure, i.e. using the \textcolor{black}{plug-in} procedure described in Section~\ref{sec:ateestimation}. This procedure will be referred to as within-trial prognostic score adjustment. \textcolor{black}{This procedure can be summarized as:}
\textcolor{black}{
\begin{enumerate}
    \item Fit a prognostic model $\widehat \rho_1(A, W) = \hat\mu (A,W)$ on the trial data using a working outcome model potentially some powerful machine learning method.  
    \item From $\widehat \rho_1$ obtain the predicted outcomes $\widehat \rho_1(a, w)$ for each participant in the study. 
    \item Use MLE to estimate $\beta = (\beta_1, \beta_2, \beta_3)$ using the regression model $Y = \beta_1A_\pm+\beta_2\hat\mu(A, W) + X\beta_3 + \mathcal{N}(0,1)$.
    \item Then follow step 2-4 of the plug-in method described in Section~\ref{sec:plug-in_with} to obtain the ATE and variance estimate.
\end{enumerate}
}

Within-trial adjustment is equivalent to TMLE with a specific submodel. To see this, consider a 1:1 randomized trial with the goal of estimating the ATE. In the TMLE procedure we first introduced we updated an initial outcome regression $\hat\mu$ by finding the MLE $\epsilon^*$ in the model $Y = \epsilon A_\pm+ \hat{\mu}(A,W) + \mathcal N (0,1)$. However, we can choose another model to maximize over as long as the two conditions we stated above are satisfied \textcolor{black}{(a common choice is a logistic regression update \cite{GrubervanderLaan+2010}). Another legitimate submodel} is the linear regression $Y = \beta_1A_\pm+\beta_2\hat\mu(A, W) + X\beta_3 + \mathcal{N}(0,1)$. The first condition is satisfied by $\beta = (0,1,0)$ and the second by taking the derivative with respect to $\beta_1$. 

The ANCOVA step of within-trial prognostic adjustment is precisely this TMLE update, letting the in-trial prognostic score $\widehat \rho_1$ play the role of the initial outcome regression $\hat\mu$. \textcolor{black}{Due to the equivalence between the robust standard error of the ANCOVA coefficient and the IF based standard error (eq. \ref{eq:var_tmle}), the TMLE (IF) standard error also coincides with the within-trial prognostic adjustment standard error from ANCOVA.}

\section{Discussion}\label{sec:discussion}

\textcolor{black}{
Since within-trial prognostic adjustment is a TMLE, it is a consistent estimator of the ATE and inherits other beneficial properties (e.g. efficiency if the prognostic score models the conditional mean well).
Within-trial prognostic adjustment is therefore a valid method for increasing efficiency while eliminating the dependency on historical data in RCTs.
However, outside of pedagogical contexts, it should simply be called TMLE instead of being re-branded as ``within-trial prognostic adjustment''.
}

\textcolor{black}{
While within-trial prognostic adjustment is \textit{a} TMLE, it is not the \textit{usual} TMLE defined in Section~\ref{sec:tmle} because it includes the covariates in the update submodel as linear predictors. In almost all cases, including these covariates in the update is redundant because the within-trial prognostic scores should already capture the linear trends from the covariates. In other words, we should expect the covariate update coefficients to be near zero. Thus including them in the update step is not only unnecessary but also has no meaningful effect. This is briefly demonstrated by a simulation study in Appendix~\ref{app:sim}.
So while it is valid to use the within-trial prognostic score adjustment, there is no reason to reinvent the wheel. 
We therefore advocate for doing the regular TMLE already available in various software packages.}

When using the standard linear adjustment for a prognostic score the prognostic model can be fully prespecified both in regards to tuning and model parameters in the statistical analysis plan since it is based on historical data. For TMLE and within-trial prognostic score adjustment we would only be able to prespecify the library of candidate models used for the outcome regression or prognostic model and not the specific model parameters. The ability to prespecify prognostic model parameters does not affect the asymptotic Type \text{I} error rate but it could be seen as a benefit when working with regulatory authorities \cite{FDA_cov, EMA_cov}.

In practice, when doing within-trial prognostic adjustment or TMLE one should fit the prognostic/outcome model in a cross-estimated way to avoid overfitting \citep{cv_tmle, dml}. This allows valid inference without additional assumptions on the regression algorithms used. One way to do this would be to use a leave-one-out estimation scheme, where the model is fitted using $n-1$ participants to get the prediction of the one participant being left out of the fitting procedure. Another way would be to do v-fold cross estimation. \textcolor{black}{Cross-fitting the prognostic score in a within-trial prognostic score adjustment is identical to cross-fitting the outcome regression in TMLE. Therefore cross-fit, within-trial prognostic adjustment is a cross-fit TMLE. Although generally recommended, the added complexity of cross-fitting does not change any of the intuition of the previous sections so we have omitted discussion of it until this point.}

One could still consider including an additional prognostic score built from \textit{historical} data to the TMLE or within-trial procedures as proposed \citet{Liao2023}. \citet{Liao2023} shows that this can smaller clinical trials lead to further efficiency gains in small trials in particular. \textcolor{black}{Specifically, we would expect to see a gain in efficiency when the historical data is similar to the new trial and when the amount of historical data is large compared to the new trial. Therefore, it is not either/or; you can combine historically-built prognostic scores with TMLE (which is essentially the within-trial procedure). The comparison of the within-trial procedure and the standard prognostic score adjustment method have been examined in simulations by \citet{within_psi}.}

\subsection*{Acknowledgement}
This research was supported by Innovation Fund Denmark (grant number 2052-00044B).

\bibliographystyle{plainnat}
\bibliography{litterature}

\appendix

\section{. \hspace{0.1cm} Determining the TMLE updating step}\label{app1}

\textcolor{black}{In this section we show that $\epsilon^*$ in \eqref{eq:eps} is the maximum likelihood solution to the estimation problem $Y = \hat\mu(A, W) + \epsilon A_\pm + \mathcal N(0,1)$. We define $U_i := Y_i - \hat{\mu}(A_i, W_i)$ and $U = (U_1, U_2, \ldots,U_n)^\top$. Then regressing $U$ on $A_\pm$ the the design matrix $\X$ has rows $x=a_{\pm, i}$. Thus, we have $\hat{\eps} = (\X^\top \X)^{-1} \X^\top u$. First notice that $(\X^\top \X)^{-1}=n^{-1}$ using $A_\pm$ instead of $A$. Therefore, we have}
\begin{align}
\begin{split}
   \textcolor{black}{\hat{\eps}} &\textcolor{black}{= n^{-1} [\underbrace{1, 1, \ldots, 1}_{n_1}, \underbrace{-1, -1, \ldots, -1}_{n_0}]^\top U}\\
    &\textcolor{black}{= n^{-1}\left(\sum_{A_i=1}Y-\sum_{A_i=1}\hat\mu(1, W) - \sum_{A_i=0}Y +\sum_{A_i=0}\hat\mu(0, W)\right)}\\
    &\textcolor{black}{= \frac{1}{n}(n_1\tilde\Psi_1 - n_0\tilde\Psi_0) - \frac{1}{n} \left(\sum_{A_i=1} \hat\mu(1, W) - \sum_{A_i=0} \hat\mu(0, W) \right).}
\end{split}
\end{align}

\section{. \hspace{0.1cm} Simulation study}\label{app:sim}

\textcolor{black}{Here we examine the differences between within-trial prognostic covariate adjustment and TMLE that may occur due to the within-trial procedure including the covariates in the update submodel as linear predictors. This is done in different data generating scenarios and across different data set sizes. We simulate data based on the data generating process described in \citet{HojbjerreFrandsen2024}, but since we are only comparing the within-trial procedure and TMLE we do not simulate historical data and therefore not distributional shifts. For completeness we will restate the data generating process and scenarios. The code is available \href{https://github.com/NNEHFD/within_trial_prog}{here}.}

\textcolor{black}{The data generating process for the current trial is represented in the structural causal model in \eqref{eq:DPG_sim}.}
\textcolor{black}{
\begin{align}\label{eq:DPG_sim}
\begin{split}
        &W_1, W_2 \sim \text{Unif}(-2, 1) \\
        &W_3 \sim \mathcal{N}(0, 3) \\
        &W_4 \sim \text{Exp}(0.8) \\
        &W_5 \sim \Gamma(5, 10) \\
        &W_6, W_7 \sim \text{Unif}(1, 2) \\
        &A \sim \text{Bern}(0.5) \\
        &Y(a)|W,U = m_a(W, U) + \mathcal{N}(0, 1.3) \\
        &Y = A Y(1) + (1-A)Y(0).
\end{split}
\end{align}}
\textcolor{black}{In the homogeneous treatment effect scenario we have,
\begin{align*}
    m_0(W, U)=& \, \, 4.1 \cdot \sin{(|W_2|)} + 1.5 \cdot I(|W_4| > 0.25) + 1.5 \cdot \sin{(|W_5|)} + 1.4 \cdot I(|W_3| > 2.5)
\end{align*}
and $ m_1(W, U)= \text{ATE} + m_0(W, U).$ In the heterogeneous treatment effect scenario we use $m_0$ and, 
\begin{align*}
    m_1(W, U) =& \, \, 4.3 \cdot \sin{(|W_2|)}^2 + 1.3 \cdot I(|W_4| > 0.25) + 4.1 \cdot I(W_2 > 0) \cdot \sin{(|W_5|)} \\
    &+ 1.6 \cdot \sin{(|W_6|)} + 1.4 \cdot I(|W_3| > 2.5).
\end{align*}
In both scenarios $\text{ATE}=0.84$.}

\textcolor{black}{We fix the current trial sample size $n$=200, and examine both the homogeneous and heterogeneous data generating scenario. We also examine the effect of varying the data set size by setting $n= 50, 60, 70, \ldots, 200, 225, 250, 275, 300, 325, 350, 375, 400$. For all scenarios, we have $N=250$ runs of the simulations. For estimation of the ATE we consider within-trial prognostic score adjustment and TMLE. For the prognostic model estimation and the initial outcome model estimation for TMLE we use the Discrete Super Learner specified as in \citet{HojbjerreFrandsen2024}.}

\begin{figure}[!ht]
    \centering
    \includegraphics[width=0.8\textwidth, trim={0mm 10mm 0mm 0mm}, clip]{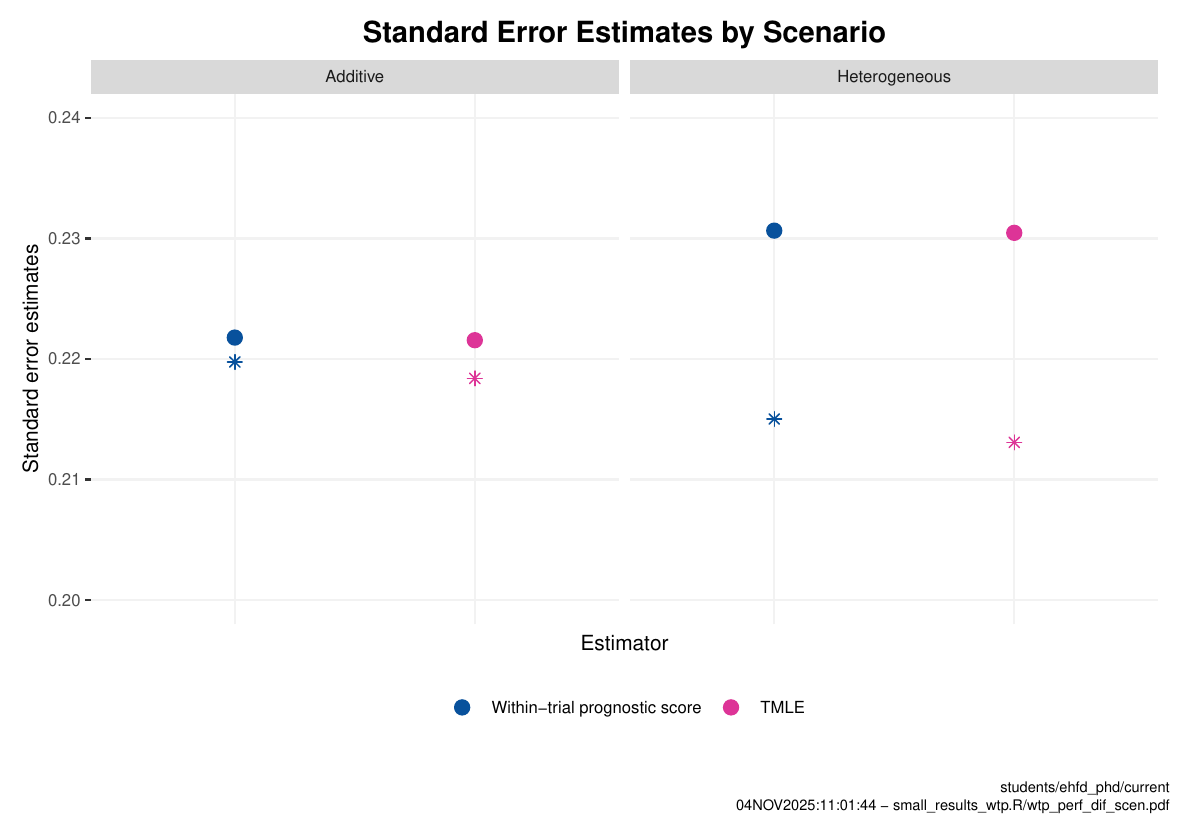}
    \caption{Standard error estimates for different scenarios. Dots represent the mean of the estimated SE, while stars represent the empirically estimated SE of the ATE estimates across the 250 simulated pairs of datasets.}
    \label{fig:perf_dif_scen}
\end{figure}

\textcolor{black}{Figure~\ref{fig:perf_dif_scen} shows the SE estimates for the estimators in the different data generating scenarios. The filled dots represent the mean of the estimated SE and the stars represents the empirically estimated SE of the ATE estimates. Unsurprisingly, the performance of linear within-trial prognostic score adjustment is similar to the TMLE across the two scenarios, but it is not identical as claimed in Section~\ref{sec:discussion} due to the covariates being included in the submodel as linear predictors.}

\begin{figure}[!ht]
    \centering
    \includegraphics[width=0.9\textwidth, trim={0mm 10mm 0mm 0mm}, clip]{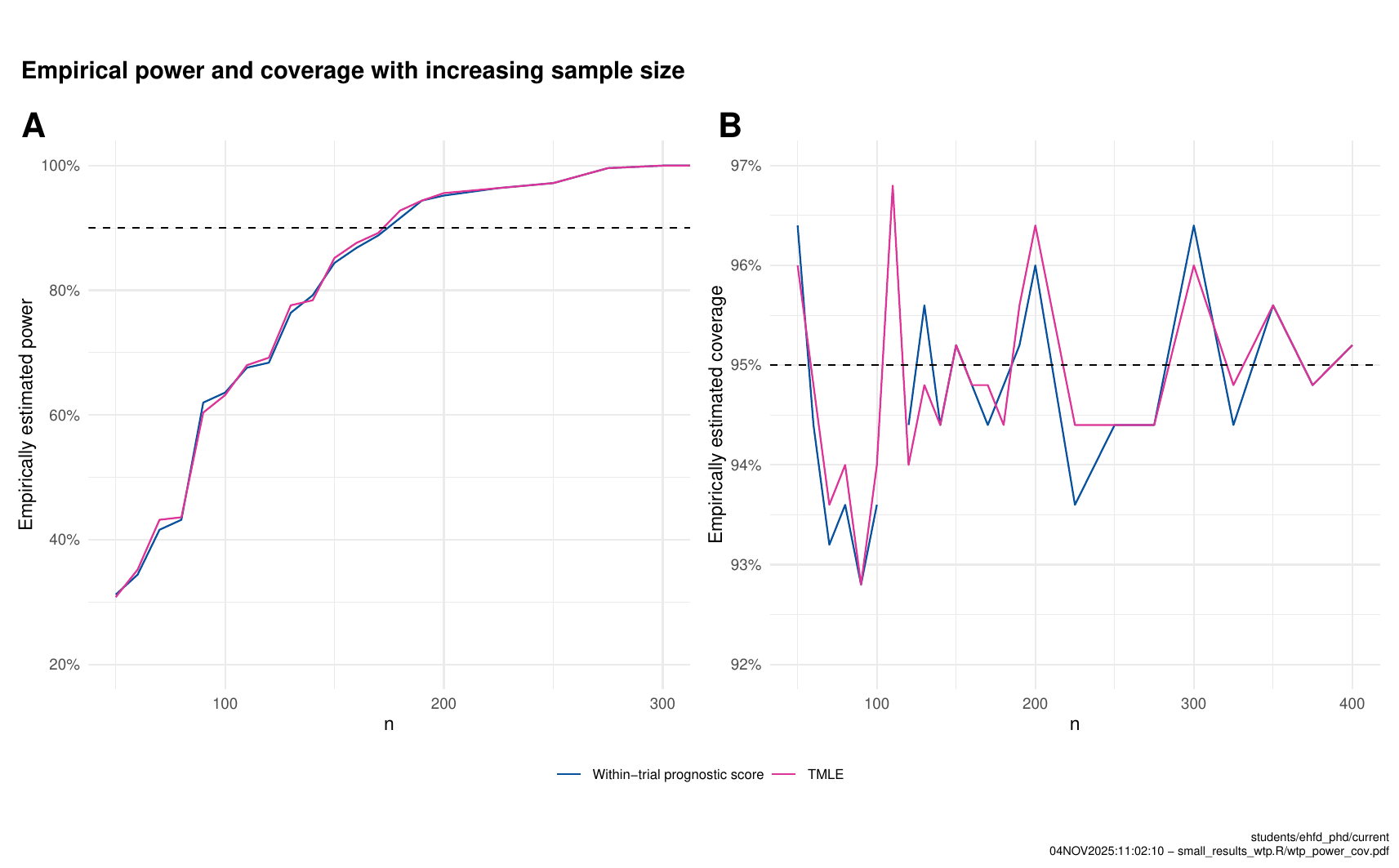}
    \caption{Empirical power and coverage with increasing sample size ($\tilde{n} = 10n$). Panel A shows the empirically estimated power for different estimators as a function of the sample size (n). Panel B displays the empirically estimated coverage for the same estimators across varying sample sizes with a significance level of 5\% (dotted black line at 95\%). }
    \label{fig:power}
\end{figure}

\textcolor{black}{In Figure~\ref{fig:power} the empirically estimated power and coverage is displayed across different data set sizes in the heterogeneous scenario. The TMLE method and within-trial prognostic score adjustment align closely for all $n$ further supporting the claim in Section~\ref{sec:discussion}. The coverage is approximately 95\% which is shown in Panel B.}

\end{document}